\documentclass[12pt,twocolumn]{article}

\usepackage[colorlinks,urlcolor=blue,linkcolor=blue,citecolor=blue]{hyperref}
\usepackage{graphicx}
\usepackage{amsmath, amssymb, amsfonts}
\usepackage{booktabs}
\usepackage{longtable}
\usepackage{adjustbox}
\usepackage{float}
\usepackage{subfig}
\usepackage{array}
\usepackage{makecell}
\usepackage[dvipsnames]{xcolor}
\usepackage{tikz}
\usetikzlibrary{fit,shapes.geometric,positioning,backgrounds,calc,arrows.meta,mindmap,shadows}
\usepackage{titlesec} 

\usepackage{authblk} 
\titleformat{\section}{\normalfont\large\bfseries}{\thesection}{1em}{}

\newcommand\blfootnote[1]{%
    \begingroup
    \renewcommand\thefootnote{}%
    \setlength{\parindent}{0pt}
    \footnote{#1}%
    \addtocounter{footnote}{-1}%
    \endgroup
}

\usepackage{mdframed}   

\usepackage[most]{tcolorbox}
\usepackage{xcolor}
\usepackage{cuted} 

\definecolor{softgray}{RGB}{245,245,245}
\definecolor{accentgray}{RGB}{200,200,200}

\newenvironment{graybox}[1][]{
  \begin{figure*}[t]
  \centering
  \begin{minipage}{1\textwidth}
  \begin{mdframed}[
    backgroundcolor=softgray,
    linecolor=accentgray,
    linewidth=1pt,
    roundcorner=6pt,
    innerleftmargin=10pt,
    innerrightmargin=10pt,
    innertopmargin=10pt,
    innerbottommargin=10pt
  ]
  \ifx\\#1\\\else{\bfseries #1}\par\smallskip\fi
  \small
}{
  \end{mdframed}
  \end{minipage}
  \end{figure*}
}
\author[1]{Fernanda Miyuki Yamada}
\author[2]{João Paulo Gois}
\author[1]{Hiroki Takahashi}

\affil[1]{The University of Electro-Communications, Chofu, Tokyo, Japan}
\affil[2]{Federal University of ABC, Santo André, São Paulo, Brazil}

\date{}

\begin{document}


\title{Toward Inclusive Avatar Design with Limb
 Differences Through Artificial Intelligence}


\twocolumn[
\vspace*{-6em}
\maketitle
\begin{abstract}
{\small As extended reality becomes more popular for social interaction and entertainment, 3D avatars must represent the full diversity of body types. Most 3D avatar systems only support normative bodies and do not accurately depict people with limb differences, amputations, or other morphological variations. This paper reviews emerging technical approaches for inclusive 3D avatar customization for this group and current guidelines that promote respectful and accurate representation. We highlight persistent challenges, including the scarcity of diverse datasets and the limitations in animation for non-normative anatomies. This paper positions artificial intelligence as a promising path to overcoming these limitations and advancing inclusive 3D avatar generation.}
\end{abstract}
\vspace*{1em}
]

\noindent \blfootnote{© 2026 IEEE. Personal use of this material is permitted. Permission from IEEE must be obtained for all other uses, in any current or future media, including reprinting/republishing this material for advertising or promotional purposes, creating new collective works, for resale or redistribution to servers or lists, or reuse of any copyrighted component of this work in other works. \\ DOI: 10.1109/MCSE.2026.3685086}

\section{Introduction}

In extended reality (XR) applications, 3D avatars mediate the user interaction with digital environments, shaping self-representation and overall experience. Despite recent advances in modeling and animation, most 3D avatar customization systems continue to default to normative bodies. This lack of morphological diversity affects mainly users with visible physical disabilities, such as limb differences, amputations, or other anatomical variations. Such users are often forced to choose between misrepresenting themselves through normative 3D avatars or forgoing authentic self-representation entirely. Consequently, the inability to create accurate self-representations in XR can reinforce feelings of exclusion present in physical spaces.

Inclusive 3D avatars offer more than just self-representation. They can impact areas such as virtual sports, rehabilitation, and education. For example, while digital games frequently simulate sports for normative bodies, there is a notable lack of virtual environments inspired by adapted sports. This gap highlights the potential of inclusive 3D avatars to increase the visibility of Paralympic athletes, challenging traditional assumptions about ability and promoting greater appreciation for the performance and dedication of athletes with disabilities.

Achieving these outcomes remains challenging due to the scarcity of 3D anatomical representations that capture diverse body types and the lack of established design standards to guide developers. Recent progress in artificial intelligence (AI) provides promising means to address these limitations. Advances in generative modeling, computer vision, and physics-based simulation have enabled accessible 3D avatar generation and improved representation of morphological diversity in virtual environments.

This paper presents an overview of emerging approaches for inclusive 3D avatar generation, with a focus on users with limb differences, amputations, and other morphological variations. Refer to the Key Terminology box for definitions of terms used throughout the paper. We advocate for digital adapted sports as a domain where such 3D avatars can have a high social and cultural impact. We emphasize the interplay between technical innovation and ethical design, reviewing existing applications and systems that address anatomical diversity, dataset considerations, and design guidelines that support diverse representation. Finally, we discuss future directions in which AI may play a critical role in enabling accessible, socially responsible XR applications.

\begin{graybox}[Key Terminology]

\begin{itemize}
\setlength{\itemsep}{4pt}
\item \textbf{Limb difference:} a congenital or acquired condition in which one or more limbs are absent or partially formed. The term refers to variations in limb presence and morphology and does not imply any specific level of functionality.

\item \textbf{Amputation:} surgical or traumatic removal of a limb or part of a limb. This term refers specifically to acquired cases and is distinct from congenital limb differences.

\item \textbf{Morphological variations:} differences in body structure, proportions, or configuration relative to the standardized human model commonly presented in medical contexts. These include, but are not limited to, limb differences and variations in skeletal morphology.

\item \textbf{Non-normative anatomy:} anatomical presentation characterized by morphological variations, referring to bodies that differ from the conventional anatomical reference models used in medical contexts.

\item \textbf{Prosthetic/Assistive device:} a wearable device that replaces or supplements a missing or impaired body part or function. It may be designed to restore lost capabilities, as in prosthetic devices, or to enhance and assist existing functions, as in assistive devices.

\end{itemize}

\end{graybox}


\section{Self-representation and Customization Systems}

Self-representation refers to how the user or player is embodied and visually represented within a digital environment. Systems that let players represent themselves, instead of only controlling a predefined protagonist, can strongly influence how interactive content is experienced. Despite this importance, many games still provide only narrow support for self-representation in practice. A large share of action and adventure titles use fixed protagonists defined by the narrative. When customization exists, it often covers only surface elements such as clothing and accessories. This pattern appears in titles such as Until Dawn, Death Stranding, and The Last of Us.

Character customization systems allow players to change the appearance of the on-screen character, instead of using a fixed protagonist. Market indicators show that customizable avatars are increasingly common in commercial games. Golando~\cite{golando2023nerf} noted that 7 of the top 13 best-selling video game titles as of February 2023, approximately 54\%, use a ``Customizable Avatar'' model rather than a fixed protagonist. Similarly, metadata from Steam, the largest PC gaming platform, identifies more than 13,550 active titles explicitly tagged with ``Character Customization'' \footnote{SteamDB, ``Character Customization'' tag. Available at: \url{https://steamdb.info/tags/}. Accessed February 9th, 2026.}. At first glance, these numbers seem to support the idea that modern games increasingly enable self-representation. However, character customization does not necessarily mean true self-representation. Many systems offer only cosmetic adjustments, with little support for body-shape editing. As a result, users may interact with a customized character that still does not reflect their identity or self-image.

The gap is larger for people with limb differences and other morphological variations. Some games feature playable characters with visible limb differences without prosthetics, such as titles in the Devil May Cry and Guilty Gear series. These examples show that non-normative body structures can be integrated into gameplay without compromising the functionality. However, such characters are predefined by the developers and offered as fixed protagonists, rather than customizable options. Therefore, these designs support inclusion at the narrative level but do not enable true self-representation for people with limb differences or other morphological variations. Moreover, a content analysis of 108 game trailers released between 2006 and 2016 found that, although disability is occasionally depicted in background non-playable characters, only 1\% of the content included playable characters with disabilities~\cite{shell2021we}.

A similar pattern appears in XR systems, where users are typically represented through 3D avatars, and embodiment plays a central role in interaction and presence. Surveys of social virtual reality platforms report that customization support is limited. Handley et al.~\cite{handley2022designing} analyzed 44 social virtual reality platforms and found that only 27.3\% of systems in their catalog offered customizable features. Options for amputations or limb differences are rarely available. A growing body of XR research addresses limb differences in rehabilitation and clinical contexts. This sustained research activity indicates that 3D avatar representations with limb differences are not a niche or isolated topic, but rather an active and expanding area of XR research with demonstrated practical applications.


\section{Guidelines in Avatar Design}

Although the importance of inclusive 3D avatars is increasingly recognized, translating this goal into practice remains challenging. Some physical disabilities are easier to incorporate within standard frameworks, particularly when they do not alter the underlying anatomy or skeletal structure of the 3D model. For instance, hearing devices, wheelchairs, or lower limb prosthetics can often be added as separate items. This discrepancy creates an unfortunate hierarchy of representation, where some disabilities are more ``visible'' in digital spaces than others.

Addressing this gap requires not only technical innovation but also a structured understanding of inclusive design practices. In this context, clear guidelines for 3D avatar design ensure that users with limb differences are represented accurately and respectfully in XR applications. These guidelines are often developed through user-centered research, where people with disabilities, designers, and accessibility experts collaborate to identify what works and what causes exclusion.

Inclusive 3D avatar guidelines emphasize giving users flexible and granular control over their virtual representation. Zhang et al.~\cite{zhang2025inclusive} suggest using continuous controls, like sliders or text prompts, so that users can describe their unique body differences naturally. This includes adjusting body features such as limb length, symmetry, and joint movement using smooth and flexible controls.

User preferences for 3D avatar representation can vary depending on social and personal considerations. Park \& Kim~\cite{park2022people} found that some users prefer to hide their disabilities to enjoy activities they cannot perform in real life and avoid stigma, while others value showing their disabilities to raise awareness and facilitate communication. Regardless of these differences, participants agreed that customization systems must offer diverse options.

Together, these studies underscore that inclusive 3D avatar customization systems have to: (1) allow users to customize the appearance of their 3D avatars to support self-representation; (2) provide granular control over the representation of physical differences and assistive devices; (3) support context-dependent representations, allowing users to represent themselves with or without prosthetics, and reflecting personal preference and situational context; and (4) integrate user-centered design and validation. 

From a technical perspective, the most challenging and important aspect is allowing users to represent themselves without prosthetics. As discussed at the beginning of this section, physical disabilities that do not require modifications to the underlying skeleton can be added relatively easily to the design of 3D avatars. Modeling absent limbs requires deviation from normative body templates, affecting not only the mesh but also the skeleton, rigging, and animation pipelines. Consequently, people with limb differences who choose not to use prosthetics or cannot use them often remain the most ``invisible'' in virtual spaces. When limb differences are represented using prosthetics, the system defaults to normative body templates, thus only simulating the appearance of a disability rather than promoting authentic representation. The stakes are high, as accurate and realistic representations require not only careful 3D modeling but also fully adaptive animation pipelines that preserve natural motion without defaulting to normative body assumptions. In the following, we analyze current 3D applications and avatar customization systems to assess whether these guidelines are implemented or overlooked in the representation of limb differences.

\section{Current Applications and Limitations}

Considering the principles of inclusive 3D avatar design outlined in the previous section, it is notable that many systems that incorporate 3D avatars with limb differences or amputations still focus primarily on representing users with prosthetics. Although prosthetic modeling is helpful in clinical contexts, its default inclusion in 3D avatar customization can hinder authentic self-representation in social settings where users may prefer to represent themselves without prosthetics.

\begin{figure}[ht!]
    \centering
     \includegraphics[width=1\linewidth]{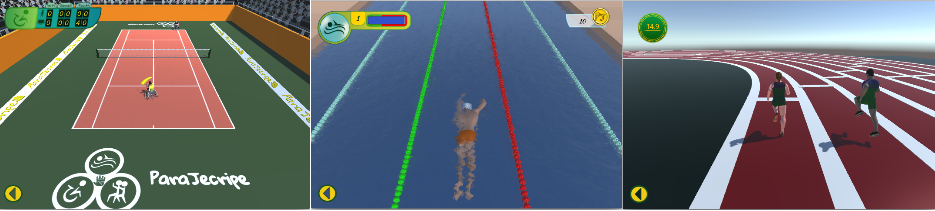}
    \caption{Parajecripe modalities for adapted sports~\cite{domingos2017production}.}
    \label{fig1}
\end{figure}

Even for applications that focus on inclusion, as in digital adapted sport games, 3D avatars with limb differences are not accurately represented and often rely on prosthetic models. Mindscape \footnote{Mindscape, 2024. Available at: \url{https://paralympics.org.uk/articles/mindscape-by-paralympicsgb-and-deloitte/}. Accessed February 9th, 2026.} allows players to create 3D avatars that use wheelchairs or prosthetic devices as visible forms of inclusion. The Pegasus Dream Tour \footnote{The Pegasus Dream Tour, 2021. Available at: \url{https://pegasus-dream.com/}. Accessed February 9th, 2026.} allows players to take part in virtual sports events using 3D avatars with prosthetics and adaptive devices. Another example is ParaJecripe~\cite{domingos2017production}, a game developed to raise awareness of adapted sports. It includes 3D avatars based on real Paralympic athletes, some of whom have limb differences and do not rely on prosthetics, but customization is limited to clothing and accessories. As shown in Figure~\ref{fig1}, ParaJecripe features avatars inspired by real Paralympic athletes, some of whom have limb differences, and are represented without prosthetics.

Only a few 3D avatar systems support people with limb differences and amputees without prosthetics, offering limited options for appearance customization. Stracke et al.~\cite{stracke2023development} present a prototype for a virtual reality simulation of lower limb amputation, where tracked movements are shown through a 3D avatar, whose template is created with the Ready Player Me system. The customization options are limited to choosing gender, height, and amputation type, which must be single-sided and either above or below the knee. The Computer Assisted Limb Assessment (CALA) system by Prahm et al.~\cite{prahm20193d}, building on MakeHuman and Blender, introduces continuous sliders for altering individual body segments such as the hands, forearms, upper arms, and shoulders. These sliders allow extreme and asymmetric modifications, ranging from thin to thick limbs, as well as complete upper limb amputation. This system partially follows the guidelines for disability representation in 3D avatars, especially regarding the recommendation for granular continuous control over the representation of physical differences. However, the process is not fully automated and still requires some level of specialized knowledge to manually adjust multiple parameters through a 3D modeling interface, such as navigating camera angles and manipulating meshes.

The literature shows that existing approaches tend to fall into two main directions: (1) medical applications, where customization focuses on accurately modeling disabilities for clinical or rehabilitative purposes, and (2) self-representation systems, which emphasize visual customization of 3D avatar appearance but often overlook the diversity of disabilities, typically relying on prosthetic additions rather than structural variation. This divide underscores the lack of methods that integrate both anatomical accuracy and expressive customization, thus not fulfilling the inclusive 3D avatar guidelines. In the following, we review some AI-based methods that aim to address these limitations.


\section{Artificial Intelligence and Morphological Editting}

Current AI-based 3D avatar generation frameworks refine models by rendering them from multiple camera viewpoints and using diffusion models to assess and improve realism. Diffusion models are a generative method that constructs data by iteratively denoising a random signal, approximating the data distribution through a reverse stochastic process learned via neural networks. Parametric models such as Skinned Multi-Person Linear Model (SMPL) and SMPL eXpressive (SMPL-X)~\cite{pavlakos2019expressive} guide this process, providing a structured representation of the human body through parameters controlling pose, shape, and sometimes facial or muscle details. Although these models ensure anatomical consistency, their design assumes normative bodies, hindering the representation of limb differences or significant asymmetries. To address this limitation, recent research has focused on modifying parametric models to handle non-normative bodies.

\begin{figure}[ht!]
    \centering
     \includegraphics[width=1\linewidth]{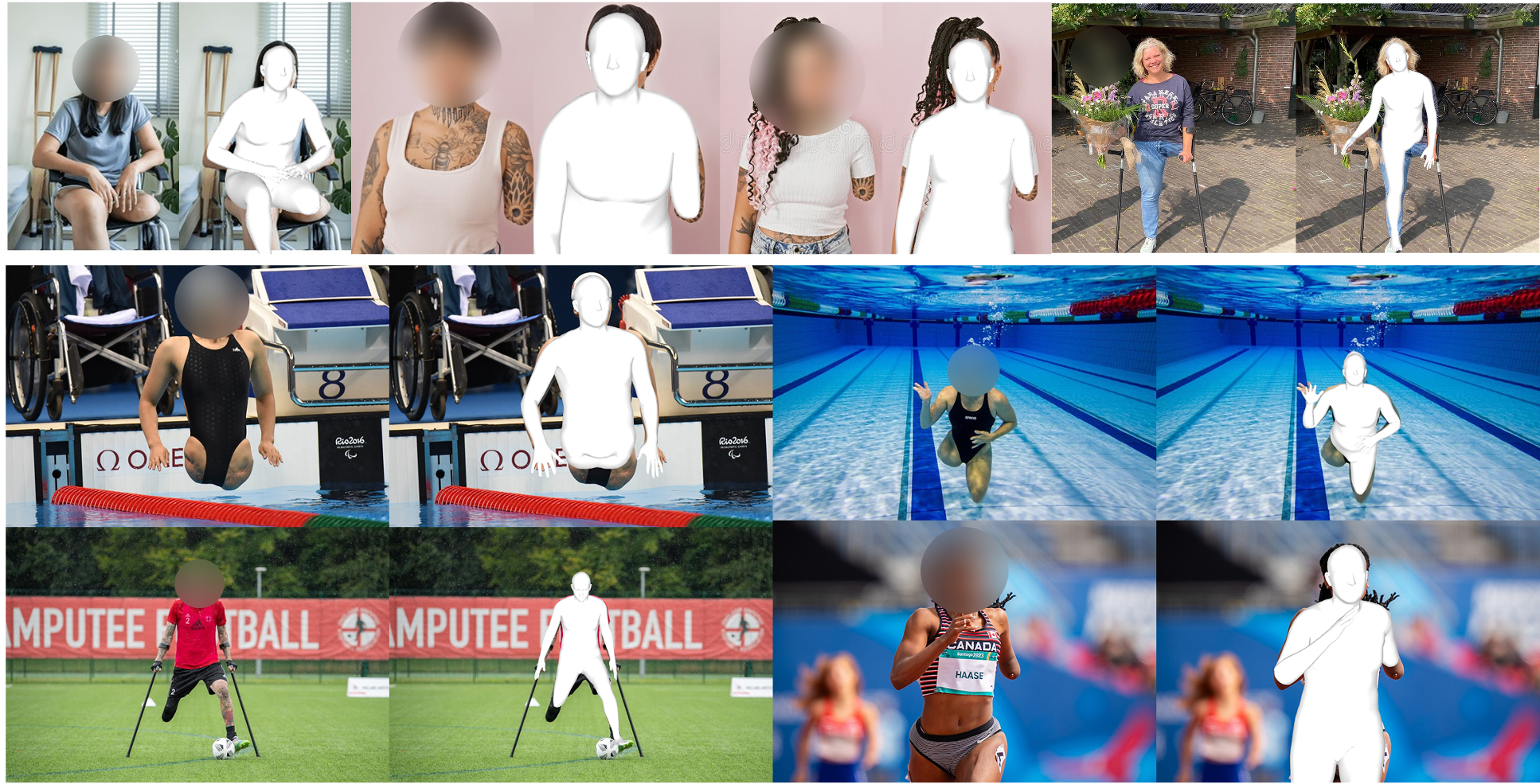}
    \caption{3D mesh recovery based on images by AJAHR~\cite{cho2025ajahr}, licensed under CC BY-SA 4.0.}
    \label{fig2}
\end{figure}

Cho et al. propose the Amputated Joint Aware 3D Human Mesh Recovery (AJAHR)~\cite{cho2025ajahr}, which reconstructs 3D human meshes from images while accounting for missing limbs. The method uses the SMPL model kinematic tree without modifying pre-trained parameters. Amputations are represented by setting the pose values of the affected joint and its descendants to zero, causing the associated vertices to collapse near the joint and simulate limb absence. A joint regressor then processes these vertices to produce anatomically consistent 3D joint positions that reflect the hierarchical effects of missing limbs. A Vision Transformer predicts the pose values from images, with separate branches for classifying the amputation status of each major limb. Global body rotation, shape, and camera position are predicted independently and combined with pose values to reconstruct the final mesh. Figure~\ref{fig2} shows the 3D meshes extracted from photographs of individuals with limb differences.

\begin{figure}[ht!]
    \centering

     \subfloat[\centering Full-body avatars in canonical pose.]{{\includegraphics[width=.45\textwidth]{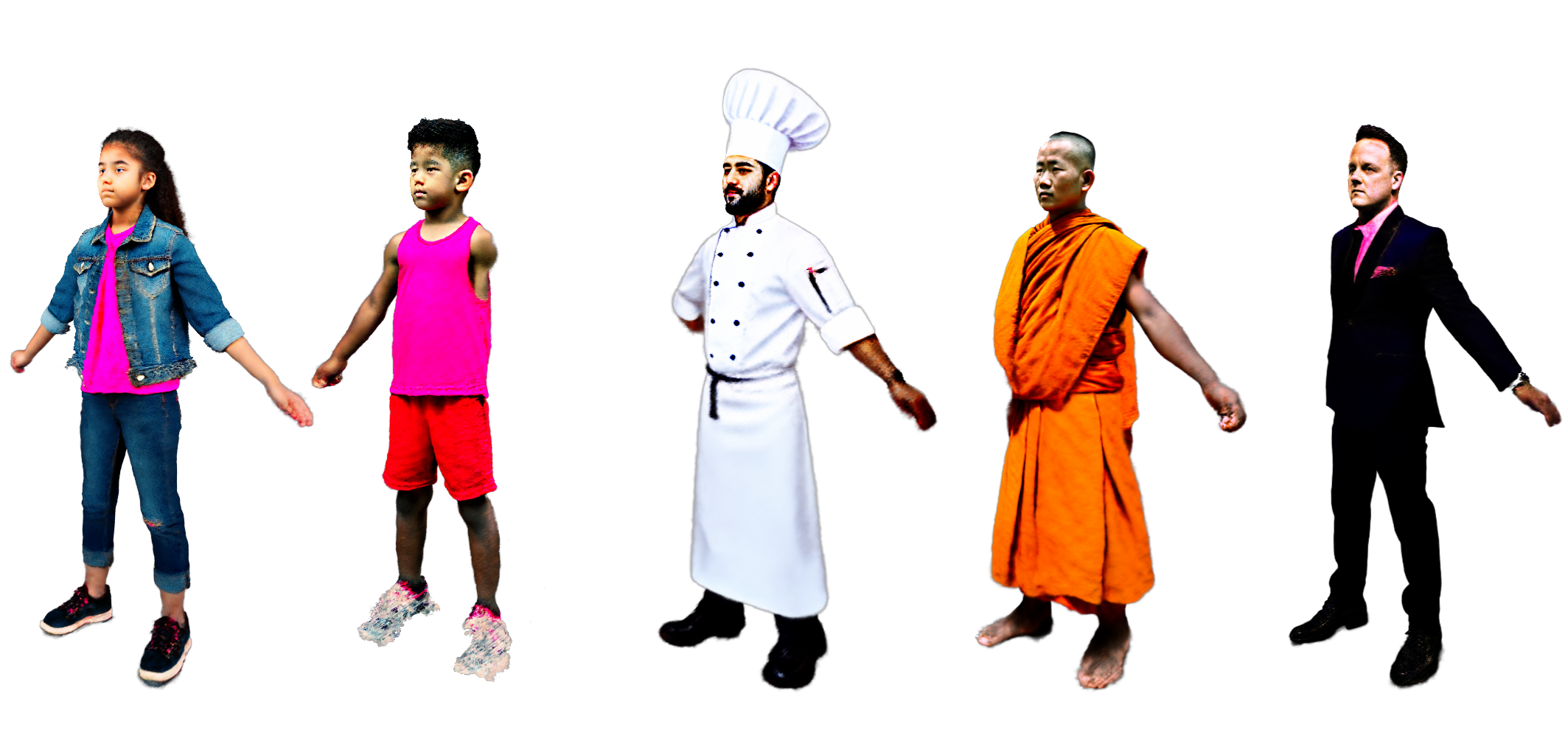} }
     \label{fig3_0}}%

    \subfloat[\centering Rendering of limb termination.]
    {\begin{minipage}{0.23\textwidth}
        \centering
        \includegraphics[height=0.8\linewidth]{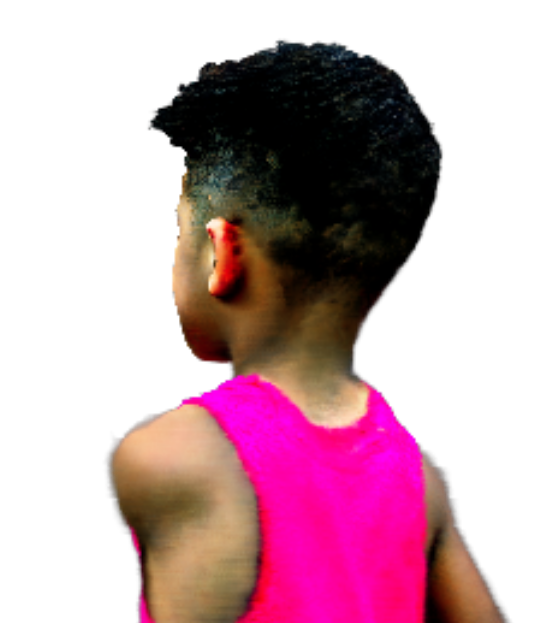} 
        \label{fig3_1}
    \end{minipage}}\hfill
    \subfloat[\centering Adaptive handling of cloth around the limb.]
    {\begin{minipage}{0.23\textwidth}
        \centering
        \includegraphics[height=0.8\linewidth]{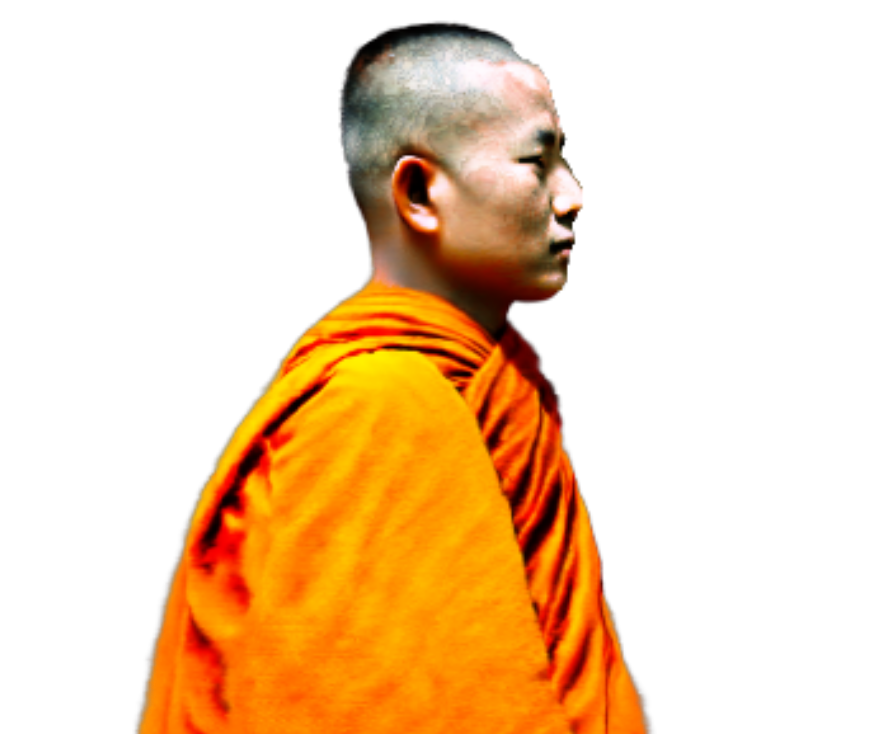} 
        \label{fig3_2}
    \end{minipage}}
    \caption{3D avatars generated by DreamAble~\cite{dreamable2026}.}
    \label{fig3}
\end{figure}

Yamada et al. propose DreamAble~\cite{dreamable2026}, a framework that generates 3D avatars with upper limb differences from natural language descriptions. The authors justify this focus by observing that upper limb prosthetics are less standardized, more context-dependent, and less commonly used than lower limb prosthetics. DreamAble allows users to specify not only their limb differences but also their appearance, clothing, and style. Instead of relying on image input or rigid-body parameters, it interprets textual descriptions of limb differences and adapts the 3D avatar accordingly. The method uses the SMPL-X model kinematic tree to remove the corresponding keypoints from an OpenPose representation of the skeleton, which is later used for skeletal guidance for a diffusion model. The authors introduce a skeleton-aware loss function that penalizes anatomical inconsistencies by comparing generated 3D avatar silhouettes against the adapted skeleton structure. Figure~\ref{fig3} illustrates results produced by DreamAble. Figure~\ref{fig3_0} shows the generated 3D avatars in a canonical pose. Figures~\ref{fig3_1} and~\ref{fig3_2} provide close-up views, highlighting limb termination and cloth positioning details, respectively.

Both methods take meaningful steps toward more inclusive 3D avatar generation, but address the inclusive guidelines in different ways. AJAHR focuses on anatomical accuracy, offering a technically precise reconstruction of limb differences through a modified joint representation. However, it remains limited to reconstructing existing human meshes and does not provide appearance customization. DreamAble allows users to describe their limb differences and appearance through natural language, but it is limited to upper limb differences. A shared limitation of both approaches is the lack of continuous control over limb length, as they only support limb differences resembling joint-level amputations. These approaches also overlook user-centered validation and community engagement.

\section{Emerging Challenges and Opportunities}

Advances in 3D modeling, animation, and AI are enabling developers to generate 3D avatars that are increasingly diverse and foster inclusive self-representation. Recent guidelines emphasize the importance of customizable 3D avatars that can accurately reflect the physical characteristics of diverse users. In the following, we discuss emerging trends and outline key open problems that must be addressed in the field.

\begin{itemize}
    \item \textbf{Clothing and appearance customization:} For individuals with limb differences, realistic clothing simulation is essential, as garments must drape naturally over the limb. Customization should extend to prosthetics, letting users modify or design assistive devices and save different versions of the same avatar. AI-based 3D avatar generation approaches, especially text-prompt-based methods, represent a promising path to provide flexible and accessible customization.

\item \textbf{Limitation in rigid parametric models}: Since parametric body models are built upon fixed skeletal joints and hierarchical connections, current AI-based systems~\cite{cho2025ajahr, dreamable2026} are limited in representing limb differences that deviate from conventional joint-level structures. Addressing this limitation requires either moving away from rigid parametric structures or extending them to handle variable limb configurations through editing mechanisms.

\item \textbf{Editing and representing congenital conditions: }Current AI-based systems are limited by their inability to model the anatomical diversity of congenital limb conditions that may occur at the ends of limbs. Modeling these complex variations requires the integration of multiple specialized classification systems, such as the Oberg-Manske-Tonkin system for the upper limbs, different types of macrodactyly, and radial longitudinal deficiency. Future research could adopt a two-step method, where the first stage employs AI-based 3D avatar generation to create a rough body model that captures limb differences up to the anatomical level where the congenital variation occurs. The second stage would then refine the granular details around the ends of the limbs through an editing tool similar to the CALA system~\cite{prahm20193d}. This refinement process could be further automated by learning from diverse datasets of limb variations.

\item \textbf{Challenges in motion capture and rigging:} Creating adaptive skeletons and rigs that accommodate diverse anatomies while preserving realistic motion remains challenging. Current motion datasets, such as Motion-X~\cite{lin2024motion}, are designed for normative body types, revealing a gap in tools for animating non-normative bodies. In particular, for lower limb differences, directly applying motions from existing datasets can produce uncanny results due to shifts in balance caused by asymmetry. Addressing this challenge will require integrating parameterized body models and motion retargeting methods that adapt to differences in limb length, joint structure, and mobility. Future research should develop motion datasets that specifically target people with limb differences. Beyond applications in AI-based 3D avatar animation, such datasets could provide benchmarks for non-normative motion and facilitate the study of compensatory strategies.

\item \textbf{Addressing visual data scarcity and diversity:} Real-world datasets on limb differences remain limited and inconsistent, making it difficult to learn from few and unbalanced samples. Advances in the generation of data can help expand data diversity and enhance model generalization. For example, Cho et al.~\cite{cho2025ajahr} introduce A3D, a synthetic dataset of varied amputee poses designed primarily for training image-based generative models, such as diffusion models, to create realistic images of non-normative bodies. Unlike motion datasets, which encode temporal dynamics and skeletal movements, datasets like A3D focus on visual representation, supporting tasks such as neural rendering for 3D avatar generation and appearance editing.

\item \textbf{Standardization, ethical guidelines, and user-centered validation:} Current research on inclusive 3D avatar generation lacks standardized protocols for evaluation, making comparisons across studies difficult. Most works rely on subjective assessments of visual quality or anatomical plausibility, overlooking measures of inclusivity and representation accuracy. To address this gap, research should incorporate feedback from users, particularly those with disabilities. This feedback can be achieved by establishing participatory design in AI-based frameworks, implementing usability testing protocols, and validating methods, while also adhering to guidelines established in the literature.

\item \textbf{Integration in XR and sports simulations:} Incorporating 3D avatars that account for diverse anatomies into deployed XR experiences presents intensified versions of the challenges discussed earlier. For instance, when developing digital adapted sport games, prosthetic variability, motion adaptation, and scarcity of data representing diverse disabilities become even more pronounced. As upper limb prosthetics vary widely in form and function, motion capture and animation should be adapted to each sport modality. Challenges in training data also arise, as diffusion models struggle in sports where limbs are obscured by water or present motion blur. If any of these aspects are not properly addressed, the digital representation cannot accurately reflect the original event, undermining the goal of inclusivity. The scarcity of media on Paralympic athletes amplifies these difficulties, exposing ethical considerations around privacy, consent, and fair representation.

\end{itemize}

\section{Conclusion}

Although current AI-based 3D avatar generation approaches face considerable technical and ethical challenges, ongoing innovations offer promising means to enable respectful and representative avatars. Recent advances in the field permit users with joint-level or limb-specific variations to customize their 3D avatars to accurately reflect their physical characteristics. However, users who do not conform to normative body structures or have complex congenital conditions may encounter limitations due to rigid kinematic tree of parametric models.

Future research should prioritize the development of adaptive editing tools that allow users to customize their avatars to reflect their preferences. Integrating principles of responsible AI development is also important to mitigate biases and prevent misrepresentation. Participatory design involving users with disabilities is essential for validating avatar systems and ensuring they meet diverse needs.

Advancing this area requires close collaboration between technical development and an understanding of human experience. A central question that remains is whether existing AI-based 3D modeling workflows designed for normative bodies can be adapted to enable genuine individual customization, or whether entirely new paradigms must be developed to achieve true inclusivity.

\section{Takeaway and Future Directions}
Recent advances in AI-based approaches show that inclusive 3D avatar design for people with limb differences is no longer a distant goal. The next priority for the community is to move beyond proof-of-concept systems and commit to delivering deployed XR applications. As a first step to achieve this goal, researchers should develop methods that jointly address anatomical accuracy and expressive customization, while also establishing standardized evaluation protocols that incorporate user-centered validation. The next step is for developers of XR applications to extend the animation and rigging pipelines to include non-normative anatomies. The final step is for XR platforms to take an active role in enhancing the visibility of individuals with limb differences, particularly in adapted sports simulations, and to partner with Paralympic athletes and disability organizations. Table~\ref{community} summarizes how each group can adopt inclusive practices and guidelines.

\begin{table*}[ht!]
\centering
\begin{tabular}{p{1.5cm} p{15cm}}
\hline

\textbf{Community} & \multicolumn{1}{c}{\textbf{Recommended Actions}}  \\
\hline
\\
Researchers \vspace{0pt} &
\vspace{-17pt}
{\small
\begin{itemize}
\setlength{\itemsep}{4pt}
\setlength{\parskip}{0pt}
\setlength{\parsep}{2pt}
\item Expand synthetic image datasets like A3D to broader disability profiles to address data scarcity
\item Extend or replace rigid 3D parametric body models to represent people with limb differences
\item Build motion capture datasets that feature people with limb differences for automatic rigging
\item Incorporate localized fine-grained modification of limbs for representing congenital conditions
\item Develop evaluation protocols that measure representation accuracy, not visual quality alone
\item Establish shared benchmarks for inclusive 3D avatar generation for comparison across methods
\item Embed participatory design with people with limb differences to ensure respectful representation
\end{itemize}
}
\\ \\
\hline
\\

Developers \vspace{0pt} &
\vspace{-17pt}
{\small
\begin{itemize}
\setlength{\itemsep}{4pt}
\setlength{\parskip}{0pt}
\setlength{\parsep}{0pt}
\item Introduce AI-based 3D avatar generation to foster self-representation of people with limb differences
\item Build interfaces to edit limb differences in 3D avatars and save versions with or without prosthetics
\item Extend animation pipelines to account for variation in limb length and mobility range
\item Ensure clothing simulations handle natural draping for 3D avatars with limb differences

\end{itemize}
}
\\ \\
\hline
\\

XR Platforms \vspace{0pt} &
\vspace{-17pt}
{\small
\begin{itemize}
\setlength{\itemsep}{4pt}
\setlength{\parskip}{0pt}
\setlength{\parsep}{0pt}
\item Develop adapted sports games as both entertainment and awareness tools for general audiences
\item Partner with Paralympic athletes and disability organizations to ensure respectful representation
\item Collaborate with researchers to fund the production of motion capture data in adapted sports
\item Govern adapted sport content with policies on consent, privacy, and fair representation
\item Treat incomplete self-representation as an accessibility failure, not a secondary compliance issue
\end{itemize}
}
\\
\\

\hline
\end{tabular}
\caption{Actions to support inclusive 3D avatar design across research, development, and XR platforms.}
\label{community}
\end{table*}

\section{Acknowledgments}
This Project was conducted with the support of the Industrial Technology Innovation Program (20023347 Development of Graph-based Intelligent Metaverse Engine for Immersive Content-sharing Service) funded by the Ministry of Trade, Industry \& Energy of the Republic of Korea.

\def\refname{References}
\bibliographystyle{IEEEtran}
\bibliography{test}




\end{document}